\documentclass[acmsmall,nonacm,screen]{acmart}
\usepackage[ngerman,main=british]{babel}
\addto\captionsngerman{}
\addto\captionsbritish{}
\setcopyright{none}
\renewcommand\footnotetextcopyrightpermission[1]{}
\hypersetup{pdftitle={A Success Model at the End of Its Life Cycle: 30 Years of Regionalised Regional Rail in Germany},
  pdfauthor={Lukas Iffländer}}
\AtEndPreamble{\hypersetup{pdflang={en-GB}}}

\begin{document}

\title{A Success Model at the End of Its Life Cycle}
\subtitle{30 Years of Regionalised Regional Rail in Germany}

\author{Lukas Iffländer}
\email{lukas.ifflaender@htw-dresden.de}
\affiliation[obeypunctuation=true]{%
  \institution{Hochschule für Technik und Wirtschaft Dresden (HTW Dresden, University of Applied Sciences)}\ and \institution{Fahrgastverband PRO BAHN e.\,V. (Federal Chair)},\ \country{Germany}}
\authorsaddresses{Contact: Prof.\,Dr. Lukas Iffländer, Professorship of Information Security, Faculty of Informatics/Mathematics, Hochschule für Technik und Wirtschaft Dresden, Friedrich-List-Platz~1, 01069~Dresden, Germany, \href{mailto:lukas.ifflaender@htw-dresden.de}{lukas.ifflaender@htw-dresden.de}}

\renewcommand{\shortauthors}{L. Iffländer}

\begin{abstract}
On 1 January 1996, Germany's federal government transferred responsibility for
regional passenger rail to the states (\textit{Länder}). Three decades
later, passenger-kilometres have doubled and modern trains run on regular
interval timetables where the former federal railway (\textit{Bundesbahn})
once thinned out services ---
regionalisation is the most successful part of the German rail reform.
Its organisational model, by contrast, has reached the end of its life
cycle: tenders now attract on average only 1.3 bids, Germany's patchwork
of transport authorities indulges in duplicate structures in every
conceivable combination, and since the cap on track-access charges was
struck down, funding no longer covers the volume of services ordered.
The first structural reform has delivered. Now regional transport needs
the second.
\end{abstract}

\maketitle
\hypersetup{pdftitle={A Success Model at the End of Its Life Cycle: 30 Years of Regionalised Regional Rail in Germany}}

\noindent\fbox{\parbox{\dimexpr\linewidth-2\fboxsep-2\fboxrule}{\small
\textbf{Online version.} An abridged version of this article was published in \textsc{Bahn-Report}, issue 5/2026. This version reproduces the manuscript as it stood on 17 July 2026 and contains the complete list of references with links; figures and developments after that date are not taken into account. The author is Professor of Information Security at HTW Dresden and Federal Chair of the German passenger association Fahrgastverband PRO BAHN; the article reflects his personal view. This English translation was prepared with AI assistance and reviewed by the author; the German original follows the English version and is the authoritative text.}}

\section{Why regionalisation happened}

The starting point of 1993 is easily forgotten: the West German Deutsche
Bundesbahn and the East German Deutsche Reichsbahn ran an annual deficit
equivalent to EUR 8 billion, the debt of the two railways reached EUR 34
billion, and the federal government's annual expenditure almost EUR 20
billion~\cite{q1}. Regional services were the easiest thing to cut:
between 1970 and 1990, federal payments to the Bundesbahn almost tripled
--- yet over the same period it cut regional passenger rail
(\textit{Schienenpersonennahverkehr}, SPNV) services by 8 per cent~\cite{q1}. Since 1955, more than 15,000 route-kilometres have disappeared
nationwide, just over a quarter of the network~\cite{q2}.

The response was radical: the Regionalisation Act
(\textit{Regionalisierungsgesetz}, RegG) declared the adequate provision
of local public transport a service of general interest
(\textit{Daseinsvorsorge}) and, with effect from 1 January 1996,
transferred responsibility for regional passenger rail to the states
(\textit{Länder}), safeguarded by the new Article 106a of the Basic Law
(\textit{Grundgesetz})~\cite{q3}. The explanatory memorandum to the bill
set out the principle that still holds today: responsibility for tasks
and for expenditure was to be `brought together in one hand' where
demand arises, complemented by competition `in order to ensure that the
most favourable offer in each case prevails'~\cite{q4}. To this end,
the states received DM 8.7 billion in 1996 and DM 12 billion annually
from 1997 onwards out of mineral oil tax revenue; in 2026 the figure is
around EUR 11.9 billion plus EUR 1.5 billion for the Deutschlandticket,
the nationwide flat-rate monthly pass for local and regional public
transport~\cite{q5}.

\section{A record that stands up to
scrutiny}\label{a-record-that-stands-up-to-scrutiny-en}

Measured against its own objectives, regionalisation is a success that
no other stage of the rail reform has achieved. Passenger traffic of DB's
regional division stood at 33.4 billion passenger-kilometres with 1.5
billion passengers in 1996~\cite{q6}; for 2024, the Federal Network
Agency (\textit{Bundesnetzagentur}, BNetzA) reported 65 billion
passenger-kilometres and 2.9 billion passengers in regional passenger
rail --- both record highs~\cite{q7}. Operating output grew from around
498 million train-kilometres (1993/94)~\cite{q8} to 708 million
train-path kilometres~\cite{q7}, i.e.\ by just over 40 per cent. The
state railway monopoly turned into a market with 125 active railway
undertakings (\textit{Eisenbahnverkehrsunternehmen}); the share of
non-DB operators in operating output rose from 9 per cent (2002)~\cite{q8} to 41 per cent~\cite{q7}. In doing so, competition unlocked
precisely the reserves the legislature had hoped for in 1993: in
first-time tenders, costs per train-kilometre fell by up to 60 per cent~\cite{q9} --- money that the transport authorities
(\textit{Aufgabenträger}) for the most part put into additional
services. Bavaria, for instance, expanded its orders from 82 to 137
million train-kilometres and today serves 92 per cent of its 1,079
stations at least hourly~\cite{q10}.

\begin{figure}[tbp]
\includegraphics[width=\linewidth]{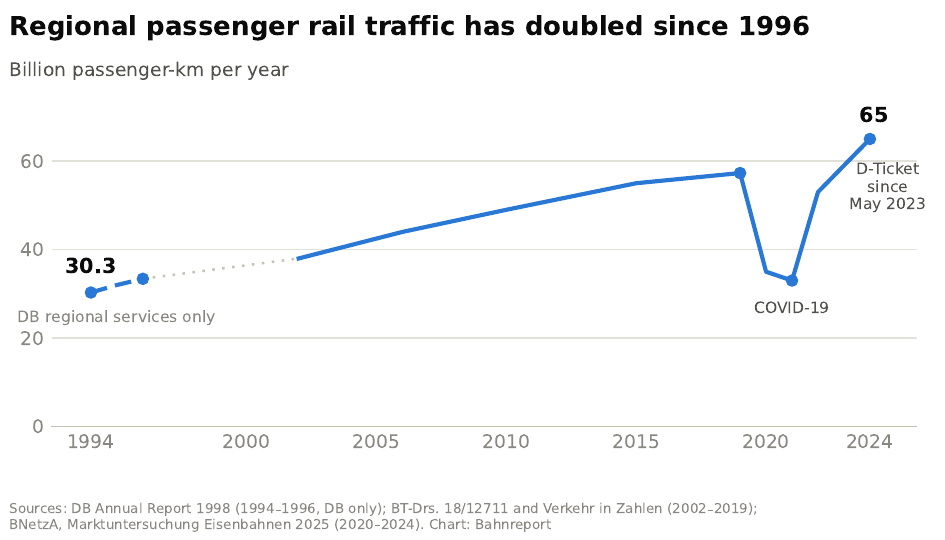}
\caption{Passenger traffic in regional passenger rail 1994--2024 in billion passenger-kilometres.}
\label{fig:verkehr-en}
\Description{Line chart of passenger traffic, rising from 30.3 billion pkm (1994) to 65 billion pkm (2024).}
\end{figure}

Reactivations also belong in this assessment --- but both sides of the
ledger must be shown: since 1994, passenger services have returned to
more than 900 route-kilometres, while over the same period services on
more than 3,600 kilometres were cancelled by the ordering authorities~\cite{q11}.

Forchheim shows how fine the line between rescue and loss was. Two
branch lines once ran from there into the surrounding countryside. The
Wiesenttalbahn to Ebermannstadt never lost its passenger services; after
regionalisation they were ordered and competitively tendered --- today
agilis runs the RB 22 hourly from early morning to late evening. The
other line, the Aischgrundbahn to Höchstadt, lost its passenger services
in 1984, twelve years before regionalisation. It was formally released
from railway use (\textit{entwidmet}) in 2008 and is now a cycle path;
the Association of German Transport Companies (VDV) rates a reactivation
as no longer feasible, `line extensively built over' --- even though
Höchstadt, with a good 13,000 inhabitants, appears on the VDV's list of
middle-order centres (\textit{Mittelzentren}) without a rail connection~\cite{q11}. The finding is uncomfortable: regionalisation froze the
network as it stood in 1996. Those who had lost their connection a few
years earlier have, as a rule, still not got it back.

\section{Design flaw one: everyone goes it
alone}\label{design-flaw-one-everyone-goes-it-alone-en}

The Regionalisation Act deliberately left organisation to the states ---
and they came up with sixteen different answers. Today, 26 transport
authority organisations order Germany's regional passenger rail services~\cite{q12}, alongside, depending on the definition, around 60 transport
and fare associations (\textit{Verkehrs- und Tarifverbünde})~\cite{q13}.
An expert report for the North Rhine-Westphalia (NRW) transport ministry
sorts the landscape into three basic types: municipal special-purpose
associations (\textit{Zweckverbände}), mixed models combining the state
and other bodies, and stand-alone state-owned companies~\cite{q14}. On
top of this come all combinations of association and ordering body: the
Rhine-Ruhr transport association VRR (\textit{Verkehrsverbund
Rhein-Ruhr}) orders services itself, the Central German transport
association MDV (\textit{Mitteldeutscher Verkehrsverbund}) is a pure
fare association, and the Bavarian Railway Company BEG
(\textit{Bayerische Eisenbahngesellschaft}) orders without any
association function. This diversity would be bearable if it worked as a
laboratory. In fact, it prevents exactly that: because each state mixes
territorial scope, contract model and responsibilities differently, it
is hardly ever possible to show rigorously which organisational form
works better --- there is no benchmark, and every reform debate starts
from scratch.

Saxony provides the textbook case. In 1996, the Free State founded the
Saxon State Transport Company (\textit{Sächsische
Landesverkehrsgesellschaft}, LVG) as a central ordering body --- a
minimal apparatus that decided on Saxony's regionalisation funds
(\textit{Regionalisierungsmittel}) and whose entire staff costs in 1998
amounted to DM 535,000~\cite{q15}. As early as 1999, the LVG was put
into liquidation, and the ordering function passed to five municipal
special-purpose associations with five transport associations --- for a
population of just over four million. In 2017, the state's Public
Transport Strategy Commission (\textit{ÖPNV-Strategiekommission})
explicitly found that this structure had proved its worth and judged
mergers to be overrated, because administrative overheads accounted for
only around 5 per cent of the funds even in the most expensive case~\cite{q16}. Seven years later, the governing coalition agreed on the
opposite: a single public transport special-purpose association and a
joint state regional transport company~\cite{q17}. The first step came
on 1 January 2026 with the merger of the special-purpose associations
Oberelbe and Oberlausitz-Niederschlesien to form the Verkehrsverbund
Ostsachsen~\cite{q18}. Whether Vogtland and Mittelsachsen will join
forces next remains open --- the goal of a single special-purpose
association ultimately affects all four remaining associations anyway.
Three decades after winding up the LVG, Saxony is thus heading back
towards a state company --- it has come full circle, via a
thirty-year detour.

North Rhine-Westphalia has recently drawn the same conclusion on a
larger scale: there, three regional rail transport authorities together
employ around 390 people for the same function; the ministry's expert
report finds that the duplicate structures leave `funds tied up that
could be used to strengthen regional passenger rail'~\cite{q14}. In May
2026, the state parliament (\textit{Landtag}) voted to merge the three
authorities into a single state-wide public-law institution
(\textit{Anstalt des öffentlichen Rechts}), Schiene.NRW, from 2027~\cite{q19}.

\section{Design flaw two: the wrong party bore the
risk}\label{design-flaw-two-the-wrong-party-bore-the-risk-en}

Tendering practice in the 2000s regularly required brand-new vehicles
at the start of operations, with lead times of two to three years
between contract award and start of service. Formally, the procurement
risk lay with the operator --- in practice, it lay with the passenger,
as two cases illustrate.

In the Augsburg electric network (\textit{E-Netz Augsburg}), DB Regio
launched the Fugger-Express in December 2008 without a single approved
new train: the Alstom class 440 multiple units it had ordered were stuck
in the approval process, and for more than a year replacement sets made
up of old coaches ran instead. When the fleet did enter service, there
was a shortfall of more than 200 seats in the Munich--Augsburg peak; it
took a PRO BAHN petition to the Bavarian state parliament, with 3,330
signatures, to secure noticeable capacity increases in 2012~\cite{q20}.
Even more costly was the approval drama surrounding the Bombardier
Talent 2: dozens of finished trains stood on sidings awaiting approval
because the Federal Railway Authority (\textit{Eisenbahn-Bundesamt}) did
not approve the class until years behind schedule. On the
Saxonia-Express Dresden--Leipzig, deployment had been planned for June
2011; regular service began in December 2012~\cite{q21}. DB and the
manufacturer settled the claim out of court in 2015 --- with a payment
in the low hundreds of millions of euros~\cite{q22}.

Individual transport authorities have long since drawn the lessons: the
Lower Saxony state transport company LNVG
(\textit{Landesnahverkehrsgesellschaft Niedersachsen}) tenders networks
that require new vehicles up to five years before the start of
operations and maintains its own vehicle pool of more than 500 vehicles,
giving all bidders equal starting conditions~\cite{q23};
Baden-Württemberg founded a state-owned public-law institution
(\textit{Landesanstalt}) for this purpose, and Bavaria offers
debt-service and redeployment guarantees. The catch: these individual
solutions never became an industry standard --- which safeguards a
bidder finds still depends today on the state in which it bids.

\section{Design flaw three: profile-raising trumps proven
service}\label{design-flaw-three-profile-raising-trumps-proven-service-en}

Transport authorities are political actors, and some procurements were
clearly also about raising their own profile. The most expensive
example came from the Rhine-Main transport association RMV
(\textit{Rhein-Main-Verkehrsverbund}): in 2019, its vehicle subsidiary
ordered 27 hydrogen trains for the Taunus network, billed as the
`world's largest fleet of fuel-cell trains', with a project volume of
around EUR 500 million over 25 years~\cite{q24}. Operations began in
December 2022 with only 11 of the 27 trains and, after repeated
emergency timetables, ended with 16 hired diesel multiple units taking
over services on the main line from early 2025 --- at the manufacturer's
expense~\cite{q25}. In April 2026, three and a half years after the
start, not a single one of the 27 trains had been fully overhauled~\cite{q26}.

The galling part: the bill was known in advance. As early as 2020, a
study by the electrical engineering association VDE put the additional
costs of fuel-cell trains compared with battery trains at up to 35 per
cent~\cite{q27}. And the LNVG, whose own hydrogen fleet ran decently in
day-to-day service, decided in 2023, after a sober market survey, not to
procure any further hydrogen trains and to switch to 102 battery
multiple units from 2029~\cite{q28}. The difference between the two
transport authorities lies not in technological courage but in the
sequence: first proof in revenue service, then the fleet decision. Where
the sequence is reversed out of a desire to be first, passengers become
the guinea pigs.

\section{The market turns}

For some years now, competition in regional passenger rail has been at a
turning point. Städtebahn Sachsen ceased operations overnight in 2019;
Abellio sought refuge in protective-shield proceedings
(\textit{Schutzschirmverfahren}) in 2021, and the Dutch state railway
sold its German business; and at the end of 2021, Keolis decided to
withdraw~\cite{q29}. In May 2025, the transport authority Nahverkehr
Westfalen-Lippe (NWL) itself finally took over eurobahn --- for a
symbolic euro~\cite{q30}. Even the remaining large private operator in
NRW, National Express, needed a contract adjustment of around EUR 200
million in early 2026; the Abellio exit had previously cost the
transport authorities around half a billion euros~\cite{q31}.

The statistics confirm the impression across the board: the average
number of bids per competitive award fell from 2.2 in 2019 to 1.3 in
2024 --- 15 award procedures attracted a mere 20 bids. At the same time,
the share of gross-cost contracts (\textit{Bruttoverträge}), in which
fare revenue goes to the ordering body, which therefore bears the
revenue risk (unlike net-cost contracts, \textit{Nettoverträge}, where
the operator keeps the fares), rose to 69 per cent~\cite{q7}. Add up
everything that has shifted, and little remains of the 1996 market
model but the façade: the transport authorities bear the revenue risk,
the vehicles increasingly belong to their pools and state public-law
institutions, and with the Baden-Württemberg state-owned SWEG, the Hesse
state-owned Hessische Landesbahn and the NWL's eurobahn, several
ordering bodies are already active in the operator market themselves.
mofair, the association of competing operators, spells out what it sees
as the consequence in no uncertain terms: if the trend continues,
`regional passenger rail would be re-nationalised and the rail reform
would have failed'~\cite{q32}.

\begin{figure}[tbp]
\includegraphics[width=\linewidth]{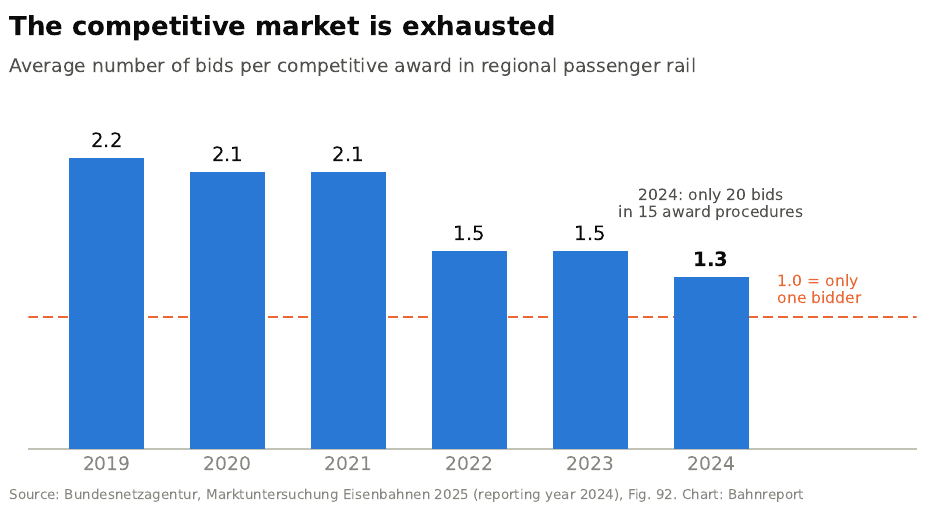}
\caption{Average number of bids per competitive award in regional passenger rail 2019--2024.}
\label{fig:bieter-en}
\Description{The number of bids per award falls from 2.2 (2019) to 1.3 (2024).}
\end{figure}

\section{The next cliff: money}\label{the-next-cliff-money-en}

The anniversary coincides with a funding crisis. In March 2026, the
Court of Justice of the European Union (CJEU) declared the track-access
charge cap (\textit{Trassenpreisbremse}) for regional passenger rail
incompatible with EU law~\cite{q33}; DB InfraGO, Deutsche Bahn's
infrastructure manager, then applied for retroactive increases of
around 10 per cent in regional rail track-access charges for 2025 and
2026~\cite{q34}. The VRR
warns of an increase of up to 23.5 per cent and puts the risk of service
cancellations in its area alone at 3.5 of 53 million train-kilometres~\cite{q35}. The states and transport authorities report an additional
funding need of EUR 14 billion for 2026 to 2031 --- and that merely to maintain
the existing level of service~\cite{q36}. For the Deutschlandticket, the
federal government has at least fixed its EUR 1.5 billion per year
until 2030~\cite{q37}; for the regionalisation funds as a whole, an
adjustment in line with actual cost trends has yet to be made. The
bitter irony is that the very reform that once freed up billions can no
longer afford its own services --- not because of the traffic, but
because of the prices charged for the federally owned infrastructure.

\begin{figure}[tbp]
\includegraphics[width=\linewidth]{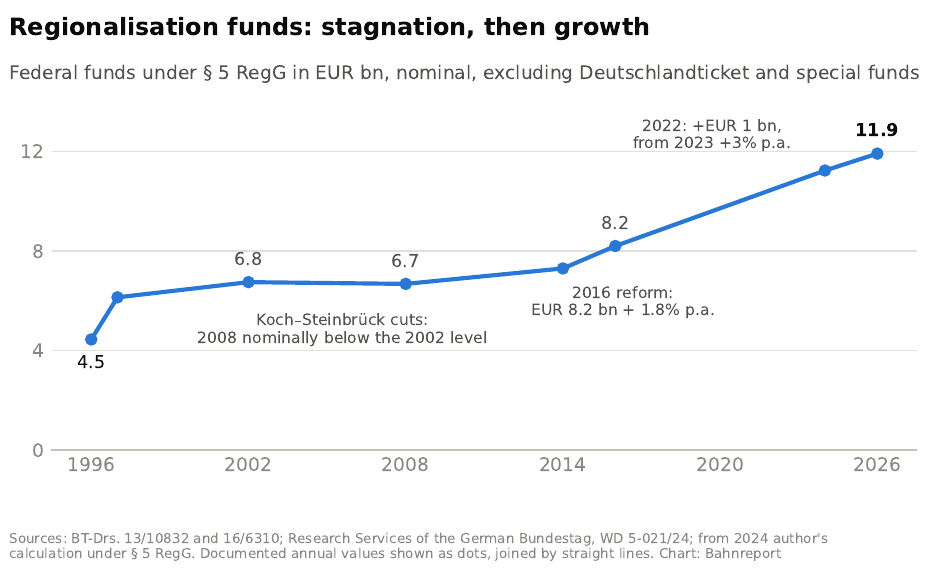}
\caption{Federal regionalisation funds 1996--2026.}
\label{fig:mittel-en}
\Description{Line chart of federal regionalisation funds in EUR bn (nominal): 4.5 (1996), around 6.7--6.8 between 2002 and 2008, 8.2 (2016), rising to 11.9 (2026).}
\end{figure}

\section{Time for the second
reform}\label{time-for-the-second-reform-en}

Thirty years of regionalisation yield four tasks.

First, consolidate structures. For subscribers, the Deutschlandticket
has delivered nationwide what was historically the fare associations'
core service: a single fare across operator boundaries. What remains are
fares for occasional travel (\textit{Bartarif}), sales and a great deal
of revenue-allocation bureaucracy (\textit{Einnahmeaufteilung}). Saxony
and NRW point the way with an association merger and Schiene.NRW; other
states should subject their structures to the same critical review of
tasks. In fairness, it must be added that administrative overheads
account for only a few per cent of the funds~\cite{q16} ---
consolidation will not balance any budget, but it concentrates
responsibility, ends duplicated work and creates the comparability that
federalism has so far denied the sector.

Second, standardise vehicles. Network-specific special requests by the
transport authorities --- from livery in state colours to the colour of
the seat covers --- turn every fleet into a short production run, make
procurement more expensive and prevent vehicles from continuing to run
cost-effectively elsewhere after the contract ends; in the view of the
non-DB operators, the only requirements that are objectively
justified, if any, are those that follow from non-uniform platform
heights~\cite{q32}. At least something is moving here: in September
2025, 27 transport authorities committed themselves in a charter to
standardising future awards for new vehicles --- with minimum series of
15 vehicles, a fixed grid of vehicle lengths and only two door widths~\cite{q40}. History gives grounds for scepticism: earlier voluntary
vehicle recommendations regularly failed because the issuing body's own
members `found good reasons for their regional special requests'~\cite{q32}. Whether the voluntary commitment becomes more than a
declaration will be decided in the coming awards --- and with ETCS
retrofitting and DB InfraGO's corridor-wide general refurbishments under
full closure (\textit{Generalsanierungen}), the price of every
go-it-alone solution rises from year to year.

Third, decide which system to adopt. Great Britain is currently
returning its railways fully to public ownership; eight out of ten
journeys there are already state-run~\cite{q38}. This is no argument
against the German approach --- London is correcting a net-cost
franchising system without public ordering bodies that had little in
common with the German model; the British White Paper itself credits
the German-style concession model with having been more successful than
Britain's own franchising~\cite{q39}. The German question is a different
one: for years, the sector has been drifting into a hybrid of half a
market and half a state railway --- gross-cost contracts,
authority-owned vehicles, ordering bodies with their own operators. Both
target models have an inner logic: a repaired competition with pools,
standard vehicles and fair risk allocation, just as much as an
integrated railway owned by a Land (\textit{Landesbahn}). What has no logic is the
permanent limbo in between, which combines the disadvantages of both
worlds. The sector must weigh the two paths soberly against each other
and decide soon --- it cannot afford endless parallel experiments,
either financially or in terms of staffing.

Fourth, make funding honest. As long as track-access and station charges
are allowed to rise faster than the regionalisation funds, every
increase merely pays for higher charges levied by the federally owned
infrastructure, not for more services. Anyone who takes seriously the
principle that the transport authorities order --- and pay for --- the
services (\textit{Bestellerprinzip}) must link the indexation to actual cost
trends or fundamentally reform the track-access charging system.

In 1993, the legislature promised to bring responsibility for regional
transport together `in one hand'~\cite{q4}. Thirty years later, it
lies with 26 transport authority organisations, around 60 associations
and a patchwork of contract models --- and yet twice as many people
travel by train as back then. That is precisely the yardstick: the first
reform proved that bold structural decisions create traffic. The second
must prove that this traffic remains affordable and manageable. Thirty
years is reason enough.

\clearpage
\begin{otherlanguage}{ngerman}
\setcounter{section}{0}
\setcounter{figure}{0}
\renewcommand{\theHsection}{de.\arabic{section}}
\renewcommand{\theHfigure}{de.\arabic{figure}}
\setlength{\emergencystretch}{1em}
\captionsetup[figure]{name={Abb.}}
\gdef\shorttitle{Erfolgsmodell am Lebenszyklusende}
\phantomsection\pdfbookmark[1]{Deutsche Originalfassung}{deutsche-originalfassung}%
\label{deutsche-originalfassung}

\begingroup
\noindent{\LARGE\sffamily\bfseries Erfolgsmodell am Lebenszyklusende\par}
\vskip2pt
\noindent{\normalsize 30 Jahre Regionalisierung des Schienenpersonennahverkehrs\par}
\vskip6pt
\noindent{\small\textsc{Deutsche Originalfassung}\par}
\endgroup
\vskip10pt

{\small\noindent\textbf{Zusammenfassung.} Zum 1. Januar 1996 übertrug der Bund die Verantwortung für den
Schienenpersonennahverkehr auf die Länder. Drei Jahrzehnte später hat
sich die Verkehrsleistung verdoppelt, fahren moderne Züge im Takt, wo
die Bundesbahn einst ausdünnte --- die Regionalisierung ist der
erfolgreichste Teil der Bahnreform. Ihr Organisationsmodell dagegen hat
das Ende seines Lebenszyklus erreicht: Auf eine Ausschreibung kommen im
Schnitt nur noch 1,3 Angebote, die Aufgabenträgerlandschaft leistet sich
Doppelstrukturen in jeder denkbaren Variante, und seit dem Wegfall der
Trassenpreisbremse trägt die Finanzierung das Bestellvolumen nicht mehr.
Die erste Strukturreform hat geliefert. Jetzt braucht der Nahverkehr die
zweite.\par}
\vskip12pt

\noindent\fbox{\parbox{\dimexpr\linewidth-2\fboxsep-2\fboxrule}{\small
\textbf{Onlinefassung.} Eine gekürzte Fassung dieses Beitrags ist im \textsc{Bahn-Report}, Ausgabe 5/2026, erschienen. Die vorliegende Fassung gibt das Manuskript mit Stand vom 17.07.2026 wieder und enthält das vollständige Quellenverzeichnis mit Links; Zahlen und Entwicklungen nach diesem Datum sind nicht berücksichtigt. Der Autor ist Professor für Informationssicherheit an der HTW Dresden und Bundesvorsitzender des Fahrgastverbands PRO BAHN; der Beitrag gibt seine persönliche Auffassung wieder.}}

\section{Warum regionalisiert wurde}

Die Ausgangslage von 1993 gerät leicht in Vergessenheit: Bundesbahn und
Reichsbahn erwirtschafteten einen Jahresfehlbetrag von umgerechnet acht
Milliarden Euro, die Verschuldung beider Bahnen erreichte 34 Milliarden
Euro, die jährlichen Aufwendungen des Bundes fast 20 Milliarden Euro~\cite{q1}. Der Nahverkehr war dabei Verschiebemasse: Zwischen 1970 und
1990 verdreifachten sich die Bundesleistungen an die Bundesbahn nahezu
--- das SPNV-Angebot kürzte sie im selben Zeitraum trotzdem um acht
Prozent~\cite{q1}. Seit 1955 verschwanden bundesweit mehr als 15.000
Streckenkilometer, gut ein Viertel des Netzes~\cite{q2}.

Die Antwort war radikal: Das Regionalisierungsgesetz erklärte die
ausreichende ÖPNV-Bedienung zur Aufgabe der Daseinsvorsorge und übertrug
die SPNV-Verantwortung zum 1. Januar 1996 den Ländern, abgesichert durch
den neuen Art. 106a GG~\cite{q3}. Die Gesetzesbegründung formulierte das
Prinzip, das bis heute trägt: Aufgaben- und Ausgabenverantwortung
sollten dort „in einer Hand zusammengeführt werden``, wo die Nachfrage
entsteht, ergänzt um Wettbewerb, „um zu gewährleisten, daß das jeweils
günstigste Angebot zum Zuge kommt``~\cite{q4}. Dafür erhielten die
Länder 1996 zunächst 8,7 Milliarden D-Mark, ab 1997 dann 12 Milliarden
D-Mark jährlich aus dem Mineralölsteueraufkommen; 2026 sind es rund 11,9
Milliarden Euro zuzüglich 1,5 Milliarden Euro für das Deutschlandticket~\cite{q5}.

\section{Eine Bilanz, die sich sehen lassen
kann}\label{eine-bilanz-die-sich-sehen-lassen-kann}

Gemessen an den eigenen Zielen ist die Regionalisierung ein Erfolg, den
keine andere Stufe der Bahnreform erreicht hat. Die Verkehrsleistung des
DB-Nahverkehrs lag 1996 bei 33,4 Milliarden Personenkilometern mit 1,5
Milliarden Reisenden~\cite{q6}; 2024 meldete die Bundesnetzagentur für
den SPNV 65 Milliarden Personenkilometer und 2,9 Milliarden Fahrgäste
--- beides Höchstwerte~\cite{q7}. Die Betriebsleistung wuchs von rund
498 Millionen Zugkilometern (1993/94)~\cite{q8} auf 708 Millionen
Trassenkilometer~\cite{q7}, also um gut 40 Prozent. Aus dem
Staatsbahn-Monopol wurde ein Markt mit 125 aktiven
Eisenbahnverkehrsunternehmen; der Anteil der Wettbewerbsbahnen an der
Betriebsleistung stieg von 9 Prozent (2002)~\cite{q8} auf 41 Prozent~\cite{q7}. Der Wettbewerb hob dabei genau die Reserven, auf die der
Gesetzgeber 1993 gehofft hatte: Bei erstmaligen Ausschreibungen sanken
die Kosten je Zugkilometer um bis zu 60 Prozent~\cite{q9} --- Geld, das
die Aufgabenträger ganz überwiegend in zusätzliche Verkehre steckten.
Bayern etwa weitete die Bestellung von 82 auf 137 Millionen Zugkilometer
aus und bedient heute 92 Prozent seiner 1.079 Stationen mindestens im
Stundentakt~\cite{q10}.

\begin{figure}[tbp]
\includegraphics[width=\linewidth]{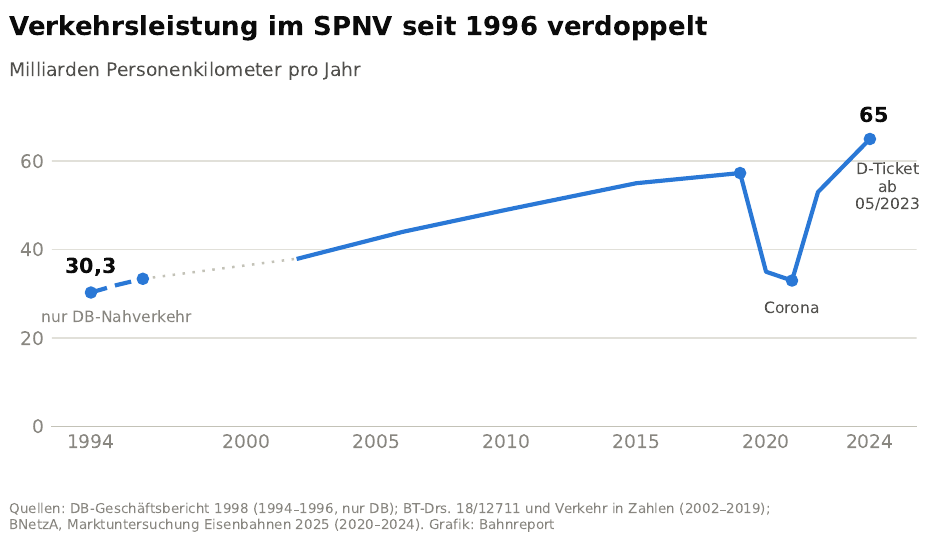}
\caption{Verkehrsleistung im SPNV 1994--2024 in Milliarden Personenkilometern.}
\label{fig:verkehr}
\Description{Liniendiagramm der Verkehrsleistung, die von 30,3 Mrd. Pkm (1994) auf 65 Mrd. Pkm (2024) steigt.}
\end{figure}

Auch die Reaktivierungsbilanz gehört in diese Rechnung, allerdings mit
beiden Seiten: Seit 1994 kehrte der Personenverkehr auf über 900
Streckenkilometer zurück --- abbestellt wurden im selben Zeitraum über
3.600 Kilometer~\cite{q11}.

Wie schmal der Grat zwischen Rettung und Verlust verlief, zeigt ein
Blick nach Forchheim. Von dort führten zwei Nebenbahnen ins Umland. Die
Wiesenttalbahn nach Ebermannstadt behielt ihren Personenverkehr
durchgehend, wurde nach der Regionalisierung bestellt und im Wettbewerb
vergeben --- heute fährt agilis die RB 22 im Stundentakt von früh bis
spät. Die parallele Aischgrundbahn nach Höchstadt verlor ihren
Personenverkehr 1984, zwölf Jahre vor der Regionalisierung. Sie wurde
2008 entwidmet und ist heute Radweg; der VDV bewertet eine Reaktivierung
als nicht mehr machbar, „Strecke stark überbaut`` --- obwohl Höchstadt
mit gut 13.000 Einwohnern in der VDV-Liste der Mittelzentren ohne
Schienenanschluss steht~\cite{q11}. Der Befund ist unbequem: Die
Regionalisierung konservierte den Bestand des Jahres 1996. Wer den
Anschluss wenige Jahre vorher verloren hatte, hat ihn in aller Regel bis
heute nicht zurück.

\section{Konstruktionsfehler eins: Jeder kocht seine
Suppe}\label{konstruktionsfehler-eins-jeder-kocht-seine-suppe}

Das Regionalisierungsgesetz überließ die Organisation bewusst den
Ländern --- und die schufen sechzehn verschiedene Antworten. Heute
bestellen 26 Aufgabenträgerorganisationen den deutschen SPNV~\cite{q12},
daneben existieren je nach Abgrenzung rund 60 Verkehrs- und
Tarifverbünde~\cite{q13}. Ein Gutachten für das NRW-Verkehrsministerium
sortiert die Landschaft in drei Grundtypen: kommunale Zweckverbände,
Mischmodelle aus Land und weiteren Trägern, einzelne
Landesgesellschaften~\cite{q14}. Dazu kommen alle Kombinationen von
Verbund und Besteller: Der VRR bestellt selbst, der MDV ist reiner
Tarifverbund, die BEG bestellt ohne jede Verbundfunktion. Diese Vielfalt
wäre verschmerzbar, wenn sie als Labor funktionierte. Tatsächlich
verhindert sie genau das: Weil jedes Land Zuschnitt, Vertragsmodell und
Zuständigkeiten anders mischt, lässt sich kaum je sauber belegen, welche
Organisationsform die bessere ist --- es fehlt der Vergleichsmaßstab,
und jede Reform beginnt argumentativ bei null.

Sachsen liefert das Lehrstück dazu. Der Freistaat gründete 1996 die
Sächsische Landesverkehrsgesellschaft (LVG) als zentralen Besteller ---
einen Kleinstapparat, der über die sächsischen Regionalisierungsmittel
entschied und dessen gesamter Personalaufwand 1998 bei 535.000 D-Mark
lag~\cite{q15}. Schon ab 1999 wurde die LVG liquidiert, die
Bestellerfunktion ging an fünf kommunale Zweckverbände mit fünf
Verkehrsverbünden --- für gut vier Millionen Einwohner. Die
ÖPNV-Strategiekommission des Landes bescheinigte dieser Struktur 2017
ausdrücklich Bewährung und hielt Fusionen für überschätzt, weil
Regiekosten selbst im teuersten Fall nur rund fünf Prozent der Mittel
ausmachten~\cite{q16}. Sieben Jahre später vereinbarte die Koalition das
Gegenteil: einen einheitlichen ÖPNV-Zweckverband und eine gemeinsame
Landesnahverkehrsgesellschaft~\cite{q17}. Den Anfang machte zum 1.
Januar 2026 die Fusion der Zweckverbände Oberelbe und
Oberlausitz-Niederschlesien zum Verkehrsverbund Ostsachsen~\cite{q18}.
Ob als Nächstes auch Vogtland und Mittelsachsen zusammenfinden, ist
offen --- das Ziel des einheitlichen Zweckverbands betrifft am Ende
ohnehin alle vier verbliebenen Verbünde. Drei Jahrzehnte nach der
Abwicklung der LVG steuert Sachsen also wieder auf eine
Landesgesellschaft zu --- der Kreis schließt sich, nur eben mit dreißig
Jahren Umweg.

Nordrhein-Westfalen zog die Konsequenz zuletzt größer auf: Dort
beschäftigen drei SPNV-Aufgabenträger zusammen rund 390 Menschen für
dieselbe Funktion; das Gutachten des Verkehrsministeriums stellt fest,
durch die Doppelstrukturen würden „Finanzmittel gebunden, die für die
Stärkung des SPNV eingesetzt werden könnten``~\cite{q14}. Der Landtag
beschloss im Mai 2026 die Bündelung in der landesweiten Anstalt
Schiene.NRW ab 2027~\cite{q19}.

\section{Konstruktionsfehler zwei: Das Risiko trug der
Falsche}\label{konstruktionsfehler-zwei-das-risiko-trug-der-falsche}

Die Ausschreibungspraxis der 2000er Jahre verlangte regelmäßig
fabrikneue Fahrzeuge zum Betriebsstart, bei Vorläufen von zwei bis drei
Jahren zwischen Zuschlag und Betriebsaufnahme. Das Beschaffungsrisiko
lag formal beim Verkehrsunternehmen --- praktisch lag es beim Fahrgast,
wie zwei Fälle exemplarisch belegen.

Im E-Netz Augsburg startete DB Regio den Fugger-Express im Dezember 2008
ohne einen einzigen zugelassenen Neuzug: Die bestellten
Alstom-Triebwagen der Baureihe 440 hingen in der Zulassung fest, über
ein Jahr fuhren stattdessen Ersatzgarnituren aus Altwagen. Als die
Flotte dann fuhr, fehlten in der Hauptverkehrszeit München--Augsburg
mehr als 200 Sitzplätze; erst eine Landtagspetition von PRO BAHN mit
3.330 Unterschriften brachte 2012 spürbare Kapazitätsnachbesserungen~\cite{q20}. Noch teurer geriet das Zulassungsdrama um den Bombardier
Talent 2: Fertig gebaute Züge standen zu Dutzenden ohne Zulassung auf
Abstellgleisen, weil das Eisenbahn-Bundesamt die Baureihe erst mit
Jahren Verspätung freigab. Auf dem Saxonia-Express Dresden--Leipzig war
der Einsatz zum Juni 2011 vorgesehen; der Regelbetrieb begann im
Dezember 2012~\cite{q21}. Den Schaden regelten DB und Hersteller 2015
außergerichtlich --- mit einer Zahlung im niedrigen dreistelligen
Millionenbereich~\cite{q22}.

Die Lehren daraus haben einzelne Aufgabenträger längst gezogen: Die LNVG
in Niedersachsen schreibt Netze mit Neufahrzeugbedarf bis zu fünf Jahre
vor Betriebsaufnahme aus und hält einen eigenen Fahrzeugpool von über
500 Fahrzeugen, der allen Bietern gleiche Startbedingungen verschafft~\cite{q23}; Baden-Württemberg gründete dafür eine Landesanstalt, Bayern
bietet Kapitaldienst- und Wiedereinsatzgarantien. Nur: Aus den
Einzellösungen wurde nie ein Sektorstandard --- welche Absicherung ein
Bieter vorfindet, hängt bis heute davon ab, in welchem Bundesland er
sich bewirbt.

\section{Konstruktionsfehler drei: Profilierung schlägt
Betriebsbewährung}\label{konstruktionsfehler-drei-profilierung-schluxe4gt-betriebsbewuxe4hrung}

Aufgabenträger sind politische Akteure, und manche Beschaffung diente
erkennbar auch der eigenen Sichtbarkeit. Das teuerste Beispiel lieferte
der RMV: 2019 bestellte seine Fahrzeugtochter 27 Wasserstoffzüge für das
Taunusnetz --- als „weltweit größte Brennstoffzellenzug-Flotte``,
Projektvolumen rund 500 Millionen Euro über 25 Jahre~\cite{q24}. Der
Betriebsstart im Dezember 2022 begann mit elf von 27 Zügen und mündete
über wiederholte Notfahrpläne schließlich darin, dass ab Anfang 2025
sechzehn angemietete Dieseltriebwagen den Verkehr der Hauptlinie
übernahmen --- auf Kosten des Herstellers~\cite{q25}. Im April 2026,
dreieinhalb Jahre nach dem Start, war noch kein einziger der 27 Züge
fertig generalüberholt~\cite{q26}.

Das Ärgerliche daran: Die Rechnung war vorher bekannt. Eine VDE-Studie
bezifferte schon 2020 die Mehrkosten von Brennstoffzellenzügen gegenüber
Batteriezügen auf bis zu 35 Prozent~\cite{q27}. Und die LNVG, deren
eigene Wasserstoffflotte im Alltag ordentlich lief, entschied 2023 nach
nüchterner Markterkundung, keine weiteren Wasserstoffzüge zu beschaffen
und ab 2029 auf 102 Akku-Triebzüge umzustellen~\cite{q28}. Der
Unterschied zwischen beiden Aufgabenträgern liegt nicht im
Technologiemut, sondern in der Reihenfolge: erst Betriebsbewährung, dann
Flottenentscheidung. Wo die Reihenfolge umgekehrt wird, weil man der
Erste sein will, wird der Fahrgast zum Testfahrer.

\section{Der Markt dreht sich}

Seit einigen Jahren erlebt der SPNV-Wettbewerb seine Wende. Die
Städtebahn Sachsen stellte 2019 den Betrieb von einem Tag auf den
anderen ein; Abellio rettete sich 2021 ins Schutzschirmverfahren, die
niederländische Staatsbahn verkaufte ihr Deutschlandgeschäft; Keolis
beschloss Ende 2021 den Rückzug~\cite{q29}. Die eurobahn übernahm im Mai
2025 schließlich der Aufgabenträger NWL selbst --- für einen
symbolischen Euro~\cite{q30}. Selbst der verbliebene große Private in
NRW, National Express, brauchte Anfang 2026 eine Vertragsanpassung von
rund 200 Millionen Euro; der Abellio-Ausstieg hatte die Aufgabenträger
zuvor etwa eine halbe Milliarde gekostet~\cite{q31}.

Die Statistik bestätigt den Eindruck flächendeckend: Die
durchschnittliche Bieterzahl je wettbewerblicher Vergabe fiel von 2,2 im
Jahr 2019 auf 1,3 im Jahr 2024 --- auf 15 Verfahren kamen ganze 20
Angebote. Parallel stieg der Anteil der Bruttoverträge, bei denen das
Erlösrisiko beim Besteller liegt, auf 69 Prozent~\cite{q7}. Rechnet man
zusammen, was sich verschoben hat, bleibt vom Marktmodell des Jahres
1996 vor allem die Fassade: Das Erlösrisiko tragen die Aufgabenträger,
die Fahrzeuge gehören zunehmend ihren Pools und Landesanstalten, und mit
SWEG, Hessischer Landesbahn und der NWL-eurobahn stehen mehrere
Besteller bereits selbst im Betreibermarkt. Der Wettbewerberverband
mofair formuliert die Konsequenz aus seiner Sicht unmissverständlich:
Setze sich die Entwicklung fort, „wäre der SPNV reverstaatlicht, die
Bahnreform gescheitert``~\cite{q32}.

\begin{figure}[tbp]
\includegraphics[width=\linewidth]{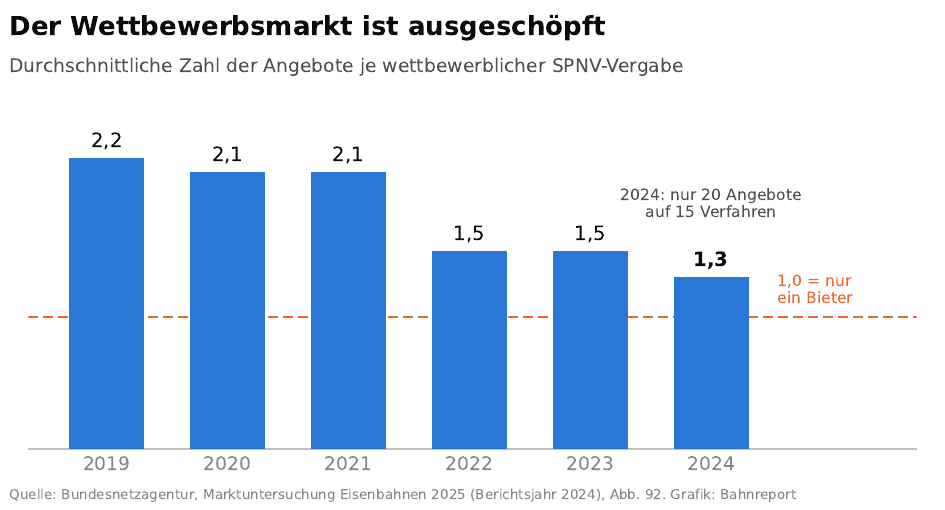}
\caption{Durchschnittliche Bieterzahl je wettbewerblicher SPNV-Vergabe 2019--2024.}
\label{fig:bieter}
\Description{Die Bieterzahl je Vergabe fällt von 2,2 (2019) auf 1,3 (2024).}
\end{figure}

\section{Die nächste Klippe: das
Geld}\label{die-nuxe4chste-klippe-das-geld}

Zum Jubiläum kommt die Finanzierungskrise. Der EuGH erklärte im März
2026 die Trassenpreisbremse für den SPNV für unvereinbar mit Europarecht~\cite{q33}; DB InfraGO beantragte daraufhin rückwirkend rund zehn
Prozent höhere SPNV-Trassenpreise für 2025 und 2026~\cite{q34}. Der VRR
warnt vor einem Anstieg um bis zu 23,5 Prozent und beziffert das
Abbestellrisiko allein in seinem Gebiet auf 3,5 von 53 Millionen
Zugkilometern~\cite{q35}. Länder und Aufgabenträger melden für 2026 bis
2031 einen Mehrbedarf von 14 Milliarden Euro --- wohlgemerkt nur, um das
bestehende Angebot zu halten~\cite{q36}. Beim Deutschlandticket hat der
Bund seine 1,5 Milliarden Euro jährlich immerhin bis 2030
festgeschrieben~\cite{q37}; für die Regionalisierungsmittel insgesamt
steht die Anpassung an die reale Kostenentwicklung aus. Die Pointe ist
bitter: Ausgerechnet die Reform, die einst Milliarden freisetzte, kann
sich ihr eigenes Angebot nicht mehr leisten --- nicht wegen des
Verkehrs, sondern wegen der Preise der bundeseigenen Infrastruktur.

\begin{figure}[tbp]
\includegraphics[width=\linewidth]{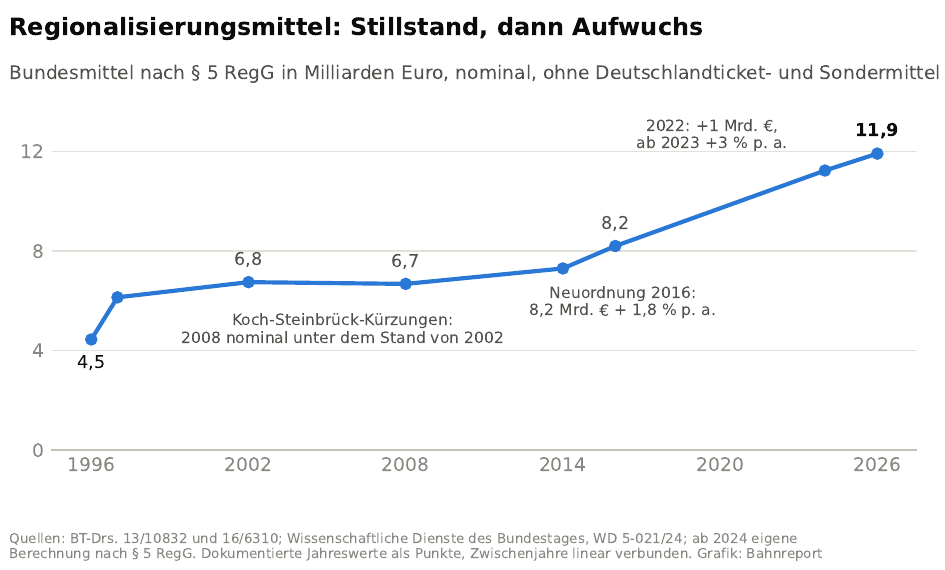}
\caption{Regionalisierungsmittel des Bundes 1996--2026.}
\label{fig:mittel}
\Description{Verlauf der Regionalisierungsmittel von 1996 bis 2026.}
\end{figure}

\section{Zeit für die zweite
Reform}\label{zeit-fuxfcr-die-zweite-reform}

Aus dreißig Jahren Regionalisierung folgen vier Aufgaben.

Erstens Strukturen konsolidieren. Das Deutschlandticket hat die
historische Kernleistung der Tarifverbünde --- einen Tarif über
Unternehmensgrenzen hinweg --- für Stammkunden bundesweit erledigt. Was
bleibt, sind Bartarif, Vertrieb und viel Einnahmeaufteilungsbürokratie.
Sachsen und NRW zeigen mit Verbundfusion und Schiene.NRW die Richtung;
andere Länder sollten ihre Strukturen derselben Aufgabenkritik
unterziehen. Ehrlicherweise gehört dazu: Regiekosten machen nur wenige
Prozent der Mittel aus~\cite{q16} --- Konsolidierung saniert keinen
Haushalt, aber sie bündelt Verantwortung, beendet Doppelarbeit und
schafft die Vergleichbarkeit, die der Föderalismus dem Sektor bisher
verweigert.

Zweitens Fahrzeuge standardisieren. Netzspezifische Sonderwünsche der
Aufgabenträger --- vom Design in Landesfarben bis zur Farbe der
Sitzbezüge --- machen aus jeder Flotte eine Kleinserie, verteuern die
Beschaffung und verhindern, dass Fahrzeuge nach Vertragsende
kostengünstig anderswo weiterfahren; objektiv gerechtfertigt sind nach
Einschätzung der Wettbewerbsbahnen allenfalls Vorgaben, die aus den
uneinheitlichen Bahnsteighöhen folgen~\cite{q32}. Immerhin bewegt sich
hier etwas: Im September 2025 verpflichteten sich 27 Aufgabenträger in
einer Charta, künftige Neufahrzeug-Vergaben zu standardisieren --- mit
Mindestserien von 15 Fahrzeugen, einem festen Raster an Fahrzeuglängen
und nur noch zwei Türbreiten~\cite{q40}. Die Skepsis liefert die
Geschichte gleich mit: Frühere freiwillige Fahrzeugempfehlungen
scheiterten regelmäßig daran, dass die eigenen Mitglieder „für ihre
regionalen Sonderwünsche gute Begründungen fanden``~\cite{q32}. Ob die
Selbstverpflichtung mehr wird als ein Bekenntnis, entscheidet sich in
den kommenden Vergaben --- und mit ETCS-Umrüstung und Generalsanierungen
steigt der Preis jedes Sonderwegs von Jahr zu Jahr.

Drittens die Systemfrage entscheiden. Großbritannien führt seine Bahnen
derzeit vollständig in öffentliche Hand zurück; acht von zehn Fahrten
fahren dort bereits staatlich~\cite{q38}. Als Argument gegen den
deutschen Weg taugt das nicht --- London korrigiert ein
Netto-Franchising ohne öffentliche Besteller, das mit dem deutschen
Modell wenig gemein hatte; das britische Weißbuch selbst bescheinigt dem
Konzessionsmodell deutscher Prägung, erfolgreicher gewesen zu sein als
das eigene Franchising~\cite{q39}. Die deutsche Frage ist eine andere:
Der Sektor driftet seit Jahren in ein Zwischending aus halbem Markt und
halber Staatsbahn --- Bruttoverträge, Behördenfahrzeuge, Besteller mit
eigenen Betreibern. Beide Zielbilder haben eine innere Logik: ein
reparierter Wettbewerb mit Pools, Standardfahrzeugen und fairer
Risikoverteilung ebenso wie die integrierte Landesbahn. Keine Logik hat
der Dauerzustand dazwischen, der die Nachteile beider Welten vereint.
Der Sektor muss beide Wege nüchtern gegeneinander abwägen und sich
zeitnah entscheiden --- ewige Parallelspielereien kann er sich weder
finanziell noch personell leisten.

Viertens die Finanzierung ehrlich machen. Solange Trassen- und
Stationspreise schneller steigen dürfen als die Regionalisierungsmittel,
kauft jede Erhöhung nur Preissteigerungen der bundeseigenen
Infrastruktur, keinen Verkehr. Wer das Bestellerprinzip ernst meint,
muss die Dynamisierung an die reale Kostenentwicklung koppeln oder die
Trassenpreissystematik grundlegend reformieren.

1993 versprach der Gesetzgeber, Verantwortung für den Nahverkehr „in
einer Hand`` zusammenzuführen~\cite{q4}. Dreißig Jahre später liegt sie
in 26 Aufgabenträgerorganisationen, rund 60 Verbünden und einem
Flickenteppich aus Vertragsmodellen --- und trotzdem fahren doppelt so
viele Menschen Bahn wie damals. Genau das ist der Maßstab: Die erste
Reform hat bewiesen, dass mutige Strukturentscheidungen Verkehr
schaffen. Die zweite muss beweisen, dass er bezahlbar und organisierbar
bleibt. Dreißig Jahre sind dafür Anlass genug.
\end{otherlanguage}

\begingroup\raggedright

\endgroup


\begin{thebibliography}{40}
\bibitem{q1}
Deutsche Bahn AG, „Bausteine und Bilanz der Bahnreform“, Berlin, 2014. \url{https://www.deutschebahn.com/resource/blob/264418/8cac7d10f5b53ef3f7ba68ef5d459a01/bilanz_bahnreform-data.pdf} (Abruf: 17.07.2026).

\bibitem{q2}
S. Gäbler, M. Krause und F. Rösel, „15 000 Kilometer Bahnstrecken weniger als vor 70 Jahren in Deutschland — Ost und West gleichermaßen betroffen“, ifo Dresden berichtet, Nr. 4/2021, 2021. \url{https://www.ifo.de/DocDL/ifoDD_21-04_03-06_Gaebler.pdf} (Abruf: 17.07.2026).

\bibitem{q3}
Gesetz zur Regionalisierung des öffentlichen Personennahverkehrs (Regionalisierungsgesetz) vom 27.12.1993 (BGBl. I S. 2378, 2395), zuletzt geändert 25.11.2025 (BGBl. 2025 I Nr. 287). \url{https://www.gesetze-im-internet.de/regg/} (Abruf: 17.07.2026).

\bibitem{q4}
Deutscher Bundestag, „Entwurf eines Gesetzes zur Neuordnung des Eisenbahnwesens“, BT-Drs. 12/4609 (neu), Begründung, 1993. \url{https://dserver.bundestag.de/btd/12/046/1204609.pdf} (Abruf: 17.07.2026).

\bibitem{q5}
Wissenschaftliche Dienste des Deutschen Bundestages, „Aufwuchs der Regionalisierungsmittel“, Sachstand WD 5 - 3000 - 021/24, 22.02.2024. \url{https://www.bundestag.de/resource/blob/996246/87f4164c551cd5db7b02568922ce59a1/WD-5-021-24-pdf.pdf} (Abruf: 17.07.2026).

\bibitem{q6}
Deutsche Bahn AG, „Geschäftsbericht 1998“, Produktionskennzahlen 1994–1998, Berlin, 1999. \url{https://ir.deutschebahn.com/fileadmin/Deutsch/vor_2000/1998_gb_dbkonzern_de.pdf} (Abruf: 17.07.2026).

\bibitem{q7}
Bundesnetzagentur, „Marktuntersuchung Eisenbahnen 2025“ (Berichtsjahr 2024), Bonn, Dezember 2025. \url{https://www.bundesnetzagentur.de/SharedDocs/Mediathek/Berichte/2025/marktuntersuchungeisenbahnen2025.pdf} (Abruf: 17.07.2026).

\bibitem{q8}
Bundesregierung, „Bericht zur Vergabepraxis im SPNV“, BT-Drs. 18/12711, 14.06.2017. \url{https://dserver.bundestag.de/btd/18/127/1812711.pdf} (Abruf: 17.07.2026).

\bibitem{q9}
mofair e. V. und Netzwerk Europäischer Eisenbahnen, „Wettbewerber-Report Eisenbahn 2019/2020“, Kapitel Personenverkehr, Oktober 2020. \url{https://mofair.de/wp-content/uploads/2020/10/WBR19-20-Kapitel-Personenverkehr_20201007.pdf} (Abruf: 17.07.2026).

\bibitem{q10}
Bayerische Eisenbahngesellschaft, „30 Jahre bayerischer SPNV: eine Erfolgsgeschichte“, Pressemitteilung, 15.06.2026. \url{https://beg.bahnland-bayern.de/de/pressemitteilungen/30-Jahre-bayerischer-SPNV-eine-Erfolgsgeschichte} (Abruf: 17.07.2026).

\bibitem{q11}
Verband Deutscher Verkehrsunternehmen, „Reaktivierung von Eisenbahnstrecken“, 4. Auflage, Oktober 2024. \url{https://www.allianz-pro-schiene.de/wp-content/uploads/2024/10/reaktivierung-von-eisenbahnstrecken_vdv_2024.pdf} (Abruf: 17.07.2026).

\bibitem{q12}
Bundesverband SchienenNahverkehr, „Unser Verband“, Stand Juli 2026. \url{https://www.schienennahverkehr.de/ueber-uns/unser-verband/} (Abruf: 17.07.2026).

\bibitem{q13}
O. Dümmler, „Erfassung und Bewertung der Arbeit von Verkehrsverbünden“, Dissertation, TU Kaiserslautern, 2015. \url{https://d-nb.info/1070819298/34} (Abruf: 17.07.2026).

\bibitem{q14}
PD — Berater der öffentlichen Hand GmbH, „Eckpunktepapier: Untersuchung von Umsetzungsoptionen für eine Strukturreform des SPNV in Nordrhein-Westfalen“, im Auftrag des MUNV NRW, 25.09.2024. \url{https://www.umwelt.nrw.de/system/files/media/document/file/munv-eckpunktepapier-spnv-strukturreform-in-nrw.pdf} (Abruf: 17.07.2026).

\bibitem{q15}
Sächsisches Staatsministerium der Finanzen, „Beteiligungsbericht 2000 des Freistaates Sachsen“, Einzeldarstellung Sächsische Landesverkehrsgesellschaft mbH i. L., Dresden, 2000. \url{https://publikationen.sachsen.de/bdb/artikel/17918/documents/24011} (Abruf: 17.07.2026).

\bibitem{q16}
ÖPNV-Strategiekommission Sachsen, „Abschlussbericht ÖPNV/SPNV in Sachsen“, Dresden, 15.12.2017. \url{https://web.archive.org/web/20220706124426/https://www.verkehr.sachsen.de/download/AbschlussberichtOEPNVStrategiekommission.pdf} (Abruf: 17.07.2026).

\bibitem{q17}
CDU Sachsen und SPD Sachsen, „Koalitionsvertrag 2024–2029“, Kapitel ÖPNV, Dresden, 04.12.2024. \url{https://www.staatsregierung.sachsen.de/download/staatsregierung/241204_Koalitionsvertrag_CDU_SPD.pdf} (Abruf: 17.07.2026).

\bibitem{q18}
Landesdirektion Sachsen, „Landesdirektion Sachsen gibt grünes Licht für neuen Zweckverband Verkehrsverbund Ostsachsen (ZVVO)“, Pressemitteilung, 08.12.2025. \url{https://www.medienservice.sachsen.de/medien/news/1093124} (Abruf: 17.07.2026).

\bibitem{q19}
Ministerium für Umwelt, Naturschutz und Verkehr NRW, „Strukturreform im Schienenpersonennahverkehr“ (Landtagsbeschluss 10. ÖPNVG-Novelle, 06.05.2026). \url{https://www.umwelt.nrw.de/strukturreform} (Abruf: 17.07.2026).

\bibitem{q20}
PRO BAHN Regionalverband Augsburg, „Aktivitäten zum Fugger-Express / Landtagspetition“, 2012. \url{https://www.pro-bahn.de/schwaben/augsburg/index.php?content=fgx_pet} (Abruf: 17.07.2026).

\bibitem{q21}
Eisenbahnjournal Zughalt.de, „Saxonia-Express: Auslieferung des E-Talent 2 hat begonnen“, 17.09.2012. \url{https://www.zughalt.de/2012/09/saxonia-express-auslieferung-des-e-talent-2-hat-begonnen/} (Abruf: 17.07.2026).

\bibitem{q22}
Der Tagesspiegel, „Streit mit Deutscher Bahn beigelegt: Bombardier zahlt für Mängel an S-Bahn-Zügen“, 21.03.2015. \url{https://www.tagesspiegel.de/wirtschaft/streit-mit-deutscher-bahn-beigelegt-bombardier-zahlt-fuer-maengel-an-s-bahn-zuegen/11536572.html} (Abruf: 17.07.2026).

\bibitem{q23}
Landesnahverkehrsgesellschaft Niedersachsen, „Fahrzeugpool“ und „Ausschreibungen und Vergabeverfahren“. \url{https://www.lnvg.de/spnv/fahrzeuge-fahrzeugpool/fahrzeugpool} (Abruf: 17.07.2026).

\bibitem{q24}
Rhein-Main-Verkehrsverbund / fahma GmbH, „RMV-Tochter fahma bestellt größte Brennstoffzellenzug-Flotte der Welt bei Alstom“, Pressemitteilung, 22.05.2019. \url{https://www.h2bz-hessen.de/Nachrichten/33568} (Abruf: 17.07.2026).

\bibitem{q25}
hessenschau, „Zu viele Pannen: Im Taunus fahren 2025 wieder Diesel- statt Wasserstoffzüge“, 15.11.2024. \url{https://www.hessenschau.de/wirtschaft/zu-viele-pannen-im-taunus-fahren-2025-wieder-diesel--statt-wasserstoffzuege-v1,wasserstoffzuege-taunus-diesel-100.html} (Abruf: 17.07.2026).

\bibitem{q26}
dpa, „Reparatur aller Wasserstoffzüge im Taunus bis Jahresende“, 28.04.2026. \url{https://www.finanznachrichten.de/nachrichten-2026-04/68318392-reparatur-aller-wasserstoffzuege-im-taunus-bis-jahresende-016.htm} (Abruf: 17.07.2026).

\bibitem{q27}
VDE, „Bewertung klimaneutraler Alternativen zu Dieseltriebzügen“, Studie, 22.07.2020; referiert nach heise online. \url{https://www.heise.de/news/Studie-Wasserstoffzuege-wesentlich-teurer-als-Batteriezuege-4850008.html} (Abruf: 17.07.2026).

\bibitem{q28}
Landesnahverkehrsgesellschaft Niedersachsen, „102 neue Akku-Triebzüge für Niedersachsen“, Pressemitteilung, 26.07.2023. \url{https://www.lnvg.de/lnvg/pressemitteilungen/artikel/102-neue-akku-triebzuege-fuer-niedersachsen-lnvg-bereitet-ausschreibung-noch-fuer-2023-vor} (Abruf: 17.07.2026).

\bibitem{q29}
Monopolkommission, „Bahn 2023: Time to GO: Endlich qualitätswirksam in den Wettbewerb!“, 9. Sektorgutachten Bahn, 04.07.2023. \url{https://www.monopolkommission.de/fileadmin/monopolkommission/Indexierte_Dateien/PDFs/Sektorgutachten/Bahn/Monopolkommission-Sektorgutachten-Bahn-9-2023-Time-to-GO-qualitaetswirksam-Wettbewerb.pdf} (Abruf: 17.07.2026).

\bibitem{q30}
Eisenbahnjournal Zughalt.de, „NWL übernimmt die Eurobahn“, 06.02.2025. \url{https://www.zughalt.de/2025/02/nwl-uebernimmt-die-eurobahn/} (Abruf: 17.07.2026).

\bibitem{q31}
Eisenbahnjournal Zughalt.de, „Vertragsanpassung bei National Express“, Februar 2026. \url{https://www.zughalt.de/2026/02/vertragsanpassung-bei-national-express/} (Abruf: 17.07.2026).

\bibitem{q32}
mofair e. V., „SPNV-Markt am Scheideweg“, Positionspapier, Juni 2024, S. 4, 13, 16–17. \url{https://mofair.de/wp-content/uploads/2024/06/240616_spnv-papier.pdf} (Abruf: 17.07.2026).

\bibitem{q33}
EuGH, Urteil vom 19.03.2026, Rs. C-770/24 (Trassenpreisbremse SPNV, \S{} 37 Abs. 2 ERegG). \url{https://curia.europa.eu/} (Abruf: 17.07.2026).

\bibitem{q34}
DB InfraGO AG, „Neue Trassenpreise für 2025 und 2026 beantragt“, Kundeninformation, 23.04.2026. \url{https://www.dbinfrago.com/web/aktuelles/kund-inneninformationen/kund-inneninformationen/2026-KW17-Trassenpreise-beantragt-13808220} (Abruf: 17.07.2026).

\bibitem{q35}
Verkehrsverbund Rhein-Ruhr, „Pressestatement Oliver Wittke zu den Auswirkungen der Abschaffung der Trassenpreisbremse“, 19.03.2026. \url{https://www.vrr.de/aktuelles/newsroom/pressestatement-oliver-wittke-vrr-vorstandssprecher-zu-den-auswirkungen-der-abschaffung-der-trassenpreisbremse-fuer-den-schienenpersonennahverkehr-durch-das-eugh-urteil-vom-19-maerz-2026/} (Abruf: 17.07.2026).

\bibitem{q36}
Bundesverband SchienenNahverkehr, „Länder und Aufgabenträger fordern Erhöhung der Regionalisierungsmittel“, Pressemitteilung, 26.06.2026. \url{https://www.schienennahverkehr.de/laender-und-aufgabentraeger-fordern-erhoehung-der-regionalisierungsmittel/} (Abruf: 17.07.2026).

\bibitem{q37}
Deutscher Bundestag, „Bundestag regelt Finanzierung des Deutschlandtickets bis 2030“ (Elftes Gesetz zur Änderung des Regionalisierungsgesetzes), 07.11.2025. \url{https://www.bundestag.de/dokumente/textarchiv/2025/kw45-de-regionalisierungsgesetz-1118560} (Abruf: 17.07.2026).

\bibitem{q38}
UK Department for Transport, „Britain's biggest train operator's services enter public ownership this weekend“, Pressemitteilung, Mai 2026. \url{https://www.gov.uk/government/news/britains-biggest-train-operators-services-enter-public-ownership-this-weekend} (Abruf: 17.07.2026).

\bibitem{q39}
UK Department for Transport, „Great British Railways: The Williams-Shapps Plan for Rail“, Weißbuch CP 423, Mai 2021. \url{https://assets.publishing.service.gov.uk/media/60cb29dde90e0743ae8c29c1/gbr-williams-shapps-plan-for-rail.pdf} (Abruf: 17.07.2026).

\bibitem{q40}
Bundesverband SchienenNahverkehr, „Charta zur Standardisierung von SPNV-Fahrzeugen“, Selbstverpflichtung der Aufgabenträger, 22.09.2025. \url{https://www.schienennahverkehr.de/wp-content/uploads/2025/09/2025-09-19_Charta_Standardisierung-der-Fahrzeuge-im-SPNV-1.pdf} (Abruf: 17.07.2026).

\end{thebibliography}
\end{document}